\documentclass[preprint,aps,prd,nofootinbib,showpacs,12pt]{revtex4-1}

\usepackage{mathrsfs,amsmath,amssymb,amsfonts}
\usepackage{graphicx,graphics,epsfig,hhline,color}

\usepackage{graphicx}
\usepackage{amssymb}
\usepackage{amsmath}
\usepackage{pifont}
\usepackage[utf8]{inputenc}

\newcommand{\be}{\begin{eqnarray}}
\newcommand{\ee}{\end{eqnarray}}
\newcommand{\beq}{\begin{eqnarray}}
\newcommand{\eeq}{\end{eqnarray}}

\newcommand{\Z}{\mathbb{Z}}

\newcommand{\expm}{e^{-\beta\left(E_i-\mu\right)}}
\newcommand{\expmm}{e^{-2\beta\left(E_i-\mu\right)}}
\newcommand{\expmmm}{e^{-3\beta\left(E_i-\mu\right)}}
\newcommand{\expp}{e^{-\beta\left(E_i+\mu\right)}}
\newcommand{\expppp}{e^{-3\beta\left(E_i+\mu\right)}}

\newcommand{\bc}{\begin{center}}
\newcommand{\ec}{\end{center}}

\newcommand{\bye}{\end{document}}

\newcommand{\dfp}{\frac{d^4p}{(2\pi)^4}}

\newcommand{\psl}{p\hspace{-1.5mm}/}

\newcommand{\ave}[1]{\langle{#1}\rangle}

\newcommand{\tr}[1]{{\rm Tr}\,[{#1}]}

\begin{document}
\title{Effects of entanglement and instanton suppression at finite temperature in a SU(2)
EPNJL model with anomaly
}

\author{M. C. Ruivo}
\altaffiliation[Corresponding author: ] {M. C. Ruivo}
\email{maria@teor.fis.uc.pt}
\affiliation{Centro de F\'{\i}sica Computacional, Department of Physics,
University of Coimbra, P-3004 - 516 Coimbra, Portugal}
\author{P. Costa}
\email{pcosta@teor.fis.uc.pt}
\affiliation{Centro de F\'{\i}sica Computacional, Department of Physics,
University of Coimbra, P-3004 - 516 Coimbra, Portugal}
\author{C. A. de Sousa}
\email{celia@teor.fis.uc.pt}
\affiliation{Centro de F\'{\i}sica Computacional, Department of Physics,
University of Coimbra, P-3004 - 516 Coimbra, Portugal}
\date{\today}

\begin{abstract}
We investigate  the phase transitions characterized by deconfinement and restoration
of chiral  and axial symmetries,
at finite temperature, in the framework of QCD inspired models.
We compare the results obtained in the SU(2) Polyakov--Nambu--Jona-Lasinio model
with anomaly and in its extended version, the Entangled Polyakov--Nambu--Jona-Lasinio model.
In the last version, four-quark vertices with entanglement between the chiral condensate and the
Polyakov loop are considered.
The thermodynamics of the phase transitions, the meson spectrum, and in particular the convergence of axial
and chiral partners, will be analyzed, as well as the topological susceptibility.
We find that an explicit temperature dependence of the coupling vertices is necessary in both models
in order to have effective restoration of the U$_A$(1) symmetry.
\end{abstract}
\maketitle


\section{Introduction}
\label{sec:introduction}

The study of matter under  extreme conditions of temperatures and/or  densities is currently
an important  field of research in physics both experimentally and
theoretically. In this limit,  quantum chromodynamics (QCD), the theory of strong
interactions, predicts that  matter becomes a plasma of deconfined quarks and gluons.
In spite of the success of many aspects of the strong
interaction physics, there are important features to clarify, such as the behavior of the
masses of the mesons and the thermodynamics of strongly interacting matter in the region
of the transition from hadronic to quark matter.

Monte Carlo simulations in lattice gauge theory provide a powerful nonperturbative first
principle approach to QCD, but   the region of large densities and low temperatures
essentially remains inaccessible to lattice simulations. This encourages  the study and
enlargement of  effective models with QCD symmetries such as
Polyakov--Nambu--Jona-Lasinio (PNJL) model.
The PNJL model is an effective model which respects important symmetries of QCD action.
It contains quarks as fundamental degrees of freedom allowing for a self-consistent description
of chiral symmetry breaking, a key feature of QCD in its low temperature and density regime.
Besides, the coupling to the Polyakov loop allows one to describe the (statistical)
confinement/deconfinement phase transition by taking into account a static gluonic
field in which quarks propagate
\cite{Meisinger:1996PLB,Fukushima:2004PLB,Ratti:2005PRD,Megias:2006PRD}.
The NJL part of the model (where a pointlike interaction mimicking frozen gluon is
introduced) can describe the chiral phase transition and its associate order parameter,
the quark condensate. On the other hand, it is known that in pure gauge theory there is a
confined/deconfined phase transition whose associate parameter is the Polyakov loop.
Quarks couple simultaneously to the chiral condensate and to the Polyakov loop,
allowing one to examine the correlation between confinement and chiral symmetry breaking
whose direct relation is not yet shown in QCD.

The PNJL model has proven to be successful in reproducing lattice data concerning QCD
thermodynamics \cite{Ratti:2005PRD}: the coupling to the Polyakov loop, resulting in a
statistical suppression of the unwanted quark contributions to the thermodynamics below the critical
temperature, allows a good description compatible with lattice results. It has been also
used to describe the full phase diagram, essentially at the mean field level.

An important query on QCD thermodynamics is the proximity or coincidence of the two phase
transitions characterized respectively by restoration of chiral symmetry and deconfinement. 
Lattice QCD results concerning this subject have been a matter of debate  and, while for 
$N_f=2+1$ flavors  these two phase transitions are reported to take place at distinct temperatures  
\cite{lattice_Tc3F}, for $N_f=2$ flavors results indicate that the two phase transitions occur 
at the same temperature \cite{lattice_Tc}, $T_C=174(3)(6)$ MeV. 
This effect could be the result of strong correlations (entanglement) between the quark 
condensate and the Polyakov field, $\Phi$. An extension of the PNJL model has been proposed 
where this entanglement is taken into account, by endowing the four-quark interaction of the model 
with a dependence on the Polyakov field. This is the so-called entangled Polyakov--Nambu--Jona-Lasinio 
(EPNJL) \cite{Sakai1,Sakai2}. In this model the two phase transitions occur at the same temperature 
and it is possible to reproduce lattice values.

A longstanding question is whether the U$_A$(1) symmetry, that is broken in the vacuum at
the quantum level by instantons, is still broken in the chiral symmetric phase.
If  the amount  of the U$_A$(1) symmetry  breaking decreases with temperature, the question
is what fraction of it remains above the critical temperature and whether and when this
symmetry is restored. It has been pointed out that there are  phenomenological consequences
for the nature of the phase transition depending on  the degree of anomaly present at the
critical temperature  \cite{UA1lattice,UA1lattice2}.
Several observables  can exhibit signals of the  restoration of axial symmetry, like the
topological susceptibility, the meson axial chiral partners, and the $\eta'$ mass.
Lattice calculations have found evidence of the decrease of the topological susceptibility
with temperature in former works \cite{Alles} as well as in recent ones
\cite{UA1JLQCD,UA1lattice,UA1lattice2}.
In the last works attention has also been given to the meson correlators of chiral and
axial partners, which should become degenerate when both symmetries are restored. 
Concerning the $\eta'$,  the decrease of its mass in medium has been predicted in several 
theoretical works \cite{veneziano,Costa1,Ruivo,kunihiro}, and recent experimental 
results \cite{eta} in Au+Au collisions are compatible with a decrease of about 200 MeV, 
which could indicate the return of the 9th ``prodigal'' Goldstone boson.

QCD inspired models  have also been used to study the restoration of axial symmetry,
in particular the NJL model \cite{chi1,Ruivo} and its extended version, the PNJL model
\cite{Costa,gatto}.  A procedure used in several works to account for the decrease of 
the topological susceptibility is to allow a temperature dependence of the anomaly coefficient 
\cite{kunihiro,chi1,Ruivo}.
In previous works, we have discussed the possible  relation between the restorations of
chiral and axial symmetries,  both in   SU(3) NJL model and PNJL models including the 't Hooft
interaction \cite{Regulariz}.
It has been found that, by allowing the presence of high momentum quarks at finite temperature
(using an infinite cutoff as explained in \cite{Regulariz})  observables related with the U$_A$(1) 
symmetry breaking vanish (the topological susceptibility as well as the mass difference between 
the meson axial partners) as a natural consequence of the effective  restoration of chiral symmetry, 
without the need of additional assumptions.
However, the situation is different in the framework of the  SU(2) NJL and PNJL models,
as we have shown in a recent work \cite{Santos}. In this case, an additional mechanism that
ensures the suppression of instantons with temperature is needed in order that the meson axial
partners become degenerate.
This was achieved by taking the anomaly coefficient as a decreasing function of the temperature.
If such a mechanism is not considered, and although the meson chiral partners degenerate and the
topological susceptibility vanishes, signals of the anomaly will remain long after the critical
temperature since the meson axial partners do not converge.

The goal of the present paper, besides enlarging the previous study  exploring different schemes
of temperature dependence of the coupling coefficients,  is to extend it to the SU(2) EPNJL model.
A comparative study will be performed and the question whether there is any relation between the
deconfinement, restoration of chiral and  axial symmetries will be analyzed.
Since restoration of axial symmetry in the PNJL model can only be achieved by
assuming a temperature dependence of coupling coefficients, we will discuss whether an explicit
temperature dependence of the vertices is also necessary in the EPNJL model for this purpose or
if the temperature dependence through the Polyakov field is sufficient.

The relevant orderlike parameters will be analyzed and special attention will be paid to
the phase structure underlying the confinement-deconfinement and chiral transitions, as well as
to thermodynamic quantities, such as the pressure, the behavior of
the  topological susceptibility and the convergence of meson axial and chiral partners.
Section II is devoted to the description of the models, Sec. III to the discussion of
the results and, finally, we summarize our findings in Sec. IV.


\section{Model and Formalism}
\label{sec:model}

We will use the SU(2) PNJL model with a 't Hooft interaction simulating the U$_A$(1) anomaly
(instanton effects).
Afterwards, several mechanisms of temperature dependence of the coupling coefficients are
included, in particular the dependence of temperature through the Polyakov loop field,
leading to entanglement of deconfinement and restoration of chiral symmetry (EPNJL model).

The PNJL Lagrangian  with explicit chiral symmetry
breaking where the quarks couple to a (spatially constant) temporal background gauge
field (represented in terms of Polyakov loops) is given by \cite{Ratti:2005PRD,Pisa1}
\begin{eqnarray}
\mathcal{L}_{PNJL} &=& \bar{q}(\,i\, {\gamma}^{\mu}\,D_\mu\,-\,\hat m)q + \mathcal{L}_1 +
\mathcal{L}_2\,-\,\mathcal{U}\left(\Phi[A],\bar\Phi[A];T\right),
\label{e1}
\end{eqnarray}
with two different interacting parts
\begin{eqnarray}
\mathcal{L}_1 &=& g_1 \Big{[}(\overline{q} q)^2 + (\bar{q} i \gamma _5 \vec{\tau} q)^2 +
(\bar{q} \vec{\tau} q)^2 + (\bar{q} i \gamma _5 q)^2 \Big{],}
\label{e1_la}
\end{eqnarray}
\begin{eqnarray}
\mathcal{L}_2 &=& g_2 \Big{[}(\bar{q} q)^2 + (\bar{q} i \gamma _5 \vec{\tau} q)^2 -
(\bar{q} \vec{\tau} q)^2 - (\bar{q} i \gamma _5 q)^2 \Big{].}
\label{e1_lb}
\end{eqnarray}
The quark fields $q = (u,d)$ are defined in Dirac and color fields, respectively
with two flavors, $N_f=2$ and three colors, $N_c=3$, the coupling coefficients   $g_1$ and
$g_2$ have  dimension $energy^{-2}$, and  $\hat{m}=\mbox{diag}(m_u,m_d)$ is the current
quark mass matrix. For simplicity we assume below that $m_u=m_d=m$. 
The original NJL model is reproduced for $g_1=g_2$.

The Lagrangian (\ref{e1}) is chiral invariant in the limit where the current quark
masses vanish. Both terms $\mathcal{L}_1$ and $\mathcal{L}_2$  are invariant upon
SU(2)$_{L}\otimes$SU(2)${_R}\otimes$U(1)-type transformations, but the $\mathcal{L}_2$
component makes the Lagrangian noncovariant upon U$_A$(1) transformations. The
$\mathcal{L}_2$ term, which can be  explicitly written in the form of a determinant
(see \cite{Santos})  and   identified as an interaction induced by instantons, according
to 't Hooft, explicitly breaks the axial symmetry even in the chiral limit.

The quarks are coupled to the gauge sector via the covariant
derivative $D^{\mu}=\partial^\mu-i A^\mu$. The strong coupling
constant $g_{Strong}$ has been absorbed in the definition of $A^\mu$:
$A^\mu(x) = g_{Strong} {\cal A}^\mu_a(x)\frac{\lambda_a}{2}$, where
${\cal A}^\mu_a$ is the SU$_c(3)$ gauge field and $\lambda_a$ are the
Gell-Mann matrices. Besides, in the Polyakov gauge and at finite temperature
$A^\mu = \delta^{\mu}_{0}A^0 = - i \delta^{\mu}_{4}A^4$. \\
The Polyakov loop $\Phi$ (the order parameter of $\Z_3$ symmetric/broken phase transition
in pure gauge) is the trace of the Polyakov line defined by
$ \Phi = \frac 1 {N_c} {\langle\langle \mathcal{P}\exp i\int_{0}^{\beta}d\tau\,
A_4\left(\vec{x},\tau\right)\ \rangle\rangle}_\beta$.

The pure gauge sector is described by an effective potential
$\mathcal{U}\left(\Phi[A],\bar\Phi[A];T\right)$ chosen to reproduce at
the mean-field level the results obtained in lattice calculations:
\begin{equation}
    \frac{\mathcal{U}\left(\Phi,\bar\Phi;T\right)}{T^4}
    =-\frac{a\left(T\right)}{2}\bar\Phi \Phi +
    b(T)\mbox{ln}[1-6\bar\Phi \Phi+4(\bar\Phi^3+ \Phi^3)-3(\bar\Phi \Phi)^2],
    \label{Ueff}
\end{equation}
where
\begin{equation}
    a\left(T\right)=a_0+a_1\left(\frac{T_0}{T}\right)+a_2\left(\frac{T_0}{T}
  \right)^2\,\mbox{ and }\,\,b(T)=b_3\left(\frac{T_0}{T}\right)^3\, .
\end{equation}

The effective potential exhibits the feature of a phase transition from color confinement
($T<T_0$, { the minimum of the effective potential being at $\Phi=0$}) to color
deconfinement ($T>T_0$, the minima of the effective potential occurring at $\Phi \neq 0$).

The parameters of the effective potential $\mathcal{U}$ are given in Table \ref{table:paramPNJL}.
These parameters have been fixed in order to reproduce the lattice
data for the expectation value of the Polyakov loop and QCD thermodynamics in the pure
gauge sector \cite{Kaczmarek:2002mc,Kaczmarek:2007pb}.

\begin{table}[t]
    \begin{center}
        \begin{tabular}{cccc}
            \hhline{|----|}
            $a_0$ & $a_1$ & $a_2$ & $b_3$ \\
            \hline
            3.51  & $-2.47$ & $15.2$ & $-1.75$  \\
            \hhline{|----|}
        \end{tabular}
    \end{center}
\caption{Parameters for the effective potential in the pure gauge sector.}
\label{table:paramPNJL}
\end{table}

The parameter $T_0$ is  the critical temperature for the deconfinement phase transition
within a pure gauge approach: it was fixed to $270$ MeV, according to lattice findings.
This choice ensures an almost exact coincidence between chiral crossover and deconfinement at
finite temperature, as observed in lattice calculations.

The Lagrangian density  (\ref{e1})  can be rewritten as
\begin{eqnarray}
{\mathcal L}_{PNJL}&=& \bar q\,(\,i\, {\gamma}^{\mu}\,D_\mu\,-\,\hat m)\,q \nonumber +
{G_s}\, [(\bar{q} q)^2 + (\bar{q} i \gamma _5 \vec{\tau} q)^2 ] +
{G_a}\,[(\bar{q} \vec{\tau} q)^2 + (\bar{q} i \gamma _5 q)^2 ]\\
&-& \mathcal{U}\left(\Phi[A],\bar\Phi[A];T\right), \label{eq:lag}
\end{eqnarray}
where, as quoted below,  $G_s=g_1+g_2$ is related to the $\pi$ and $\sigma$ mesons and
$G_a=g_1-g_2$ to $\eta$ and $a_0$ mesons.

The PNJL grand canonical  potential density in the SU$_f$(2) sector can be written as
\cite{Ratti:2005PRD,Hansen:2007PRD}
\begin{widetext}
\be \Omega(\Phi, \bar\Phi, M ; T, \mu) &=&{\cal U}\left(\Phi,\bar{\Phi},T\right)
+4G_{_{s}} N_f\left\langle\bar{q_{i}}q_{i}\right\rangle^2
- 2 N_c\,N_f \int_\Lambda\frac{\mathrm{d}^3p}{\left(2\pi\right)3}\,{E_i}\nonumber \\
&-& 2N_f\,T\int\,\frac{\mathrm{d}^3p}{\left(2\pi\right)^3}\,\left( z^+_\Phi(E_i) +
z^-_\Phi(E_i) \right) , \label{omega}
\ee
\end{widetext}
where  $E_i$ is the quasiparticle energy for the quark $i$:
$E_{i}=\sqrt{\mathbf{p}^{2}+M_{i}^{2}}$, and  $z^+_\Phi$ and $z^-_\Phi$ are the partition
function densities.

The explicit expression of $z^+_\Phi$ and $z^-_\Phi$ are given by:
\begin{align}
z^+_\Phi(E_i)
&\equiv& \mathrm{Tr}_c\ln\left[1+ L^\dagger \expp\right]=\ln\left\{1+3\left(\bar\Phi
+\Phi \expp \right)\expp+\expppp \right\}, \label{eq:termo1}
\\
z^-_\Phi(E_i) &\equiv& \mathrm{Tr}_c\ln\left[1+ L \expm \right]= \ln\left\{ 1 + 3\left(
\Phi + \bar\Phi \expm \right) \expm
 + \expmmm \right\}.
\label{eq:termo2}
\end{align}
A word is in order to describe the role of the Polyakov loop in the present model. Almost
all physical consequences of the coupling of quarks to the background gauge field stem
from the fact that in the expression of $z_\Phi$,  $\Phi$ or $\bar\Phi$ appear only as a
factor of the one- or two-quarks (or antiquarks) Boltzmann factor, for example $\expm$
and $\expmm$. Hence, when $\Phi,\bar\Phi \rightarrow 0$ (signaling what we designate as
the ``confined phase'') only $\expmmm$ remains in the expression of the grand canonical
potential, leading to a thermal bath with a small quark density. On the contrary,
$\Phi,\bar\Phi \rightarrow 1$ (in the ``deconfined phase'') gives a thermal bath with all
1-, 2- and 3-particle contributions and a significant quark density {\cite{Rossner:2007PRD,Hansen:2007PRD}.

This formalism, presented here for completeness in the grand canonical approach, will be
employed in the present work with $\mu=0$. This condition implies $\Phi=\bar\Phi$.


We can redefine the coupling constants such as the  set ($g_1, g_2$) or ($G_s, G_a$) will
be replaced by ($ G,\, c$) in the following parametrization:
\begin{equation}
G_s= g_1 + g_2 = G, \,\,\,\,\,\,\,\,\, G_a= g_1 - g_2 = G\, (1\,-\, 2\,c),
\label{redef}
\end{equation}
where  $c \,\in \,\{0,1\} $ \cite{buballa2,Brauner:2009gu} is a parameter that now
specifies the degree  of  U$_A$(1) symmetry breaking. In the present work we take $c =0.2$,
a  value which gives an adequate degree of anomaly in the vacuum \cite{Santos},
as will be shown in the sequel by the choice of model parameters.
Notice that $g_1=  G\,(1- c)$ is the coupling constant of the four-quark vertex
associated with chiral symmetry effects, while $g_2= c \,G$ is the anomaly coefficient
and is, for the present choice, $25\%$ of $g_1$.

As it is well known, in pure gauge theory, the Polyakov potential induces a first-order
phase transition at $T=T_0$. The PNJL model with the  original $T_0 = 270$ MeV, that
reproduces pure gauge lattice QCD data, yields to
a small difference between chiral and deconfinement transition temperatures.
This value is however significant when we rescale $T_0$ to 210 MeV, derived by 
renormalization-group considerations \cite{Schaefer}  so as to reproduce
the lattice QCD result, $T_d=177$ MeV  for the deconfinement
transition temperature as will be seen in the  next section. Consequently, the PNJL result
is not consistent with lattice QCD data for the transition temperatures.
This indicates that the entanglement between chiral and deconfinement transitions is weak
in this model. Following the suggestions of other authors \cite{Sakai1,Sakai2}, we implement
an explicit dependence of $G$ on $\Phi$ assuming the following form:
\begin{equation}
G(\Phi) = G\left[1\,-\,\alpha_1\,\Phi \bar\Phi\,-\,\alpha_2\,( \Phi^3\,+\,\bar\Phi^3) \right],
\label{gphi}
\end{equation}
which respect chiral, $P$, $C$ and the extended $\Z_3$ symmetries. 
We use the parameters of reference Ref. \cite{Sakai1}, where they have
been fixed to reproduce the available lattice QCD data \cite{lattice_Tc,Philipsen}.
This leads to the values $\alpha_1=\alpha_2=0.2$ and $T_0 = 170$ MeV.

A standard calculation leads straightforwardly to the gap equation and to the meson
propagators.
The following gap equations are obtained:
\begin{equation}
M_{i}=m_{i}-4G\left\langle\bar{q}q \right\rangle_{i}, \label{gap2}
\end{equation}
where one identifies $i=u,d$ and $M_{i}$ as the constituent quark mass. The quark
condensates are determined by
\begin{equation}
\left\langle\bar{q}q \right\rangle_{i}=-i\mbox{Tr}\frac{1}{\hat{p}-M_{i}}=-i\mbox{Tr}
S_{i}(p),
\label{Eq:gap}
\end{equation}
being  $S_i(p) = (\psl - M_i+i\varepsilon)^{-1}$ the propagator of quarks.

The mass spectra of the mesons is obtained by the analysis of the pole structure of the
meson propagator, given by
\begin{equation}
   1 - 4 G_{s,a} \,\Pi_{\cal M}(q^2=M_{\cal M}^2) = 0,
\label{rpamass}
\end{equation}
where
\begin{equation}
   \Pi_{\cal M}(q^2) \;=\; i \int \dfp \tr{{\cal O}_{\cal M}\,S(p+q)\,{\cal O}_{\cal M}\,S(p)}
   \label{pol}
\end{equation}
is the polarization operator for the quark-antiquark system regarding  the channel with
quantum numbers $\{{\cal M}\}$ in the mesonic sector. As mentioned above, $G_{s}$ is
related  to $\pi$ and $\sigma$ mesons and $G_{a}$ to $\eta$ and $a_0$ mesons.

The topological susceptibility, $\chi$, is an essential parameter for the study of the
problem of breaking and restoration of the U$_A$(1) symmetry.
The topological susceptibility is defined as:
\begin{equation}
 \chi=\int{\rm d}^4x\; \langle 0 |TQ(x)Q(0)|0 \rangle_{\rm c},
\end{equation}
\label{chi}
where $c$ means connected diagrams, $T$ the time order operator and $Q(x)$ is of the form
\begin{eqnarray}
Q(x)= 2 g_2  \Big{[} \mbox{det} \big{[}(\bar{q} (1 - \gamma_5) q - \mbox{det} \big{[}(\bar{q} (1 + \gamma_5)q \Big{]}.
\label{e04_lb}
\end{eqnarray}
Taking only into account the connected diagrams, and only the terms of order $1/N_c$,
following a similar approach to \cite{chi1}, we arrive at the following expression for
the topological susceptibility (see \cite{Santos} for details):
\begin{equation}
 \chi = 4 N_f\,  g_2^2\,\langle \bar q q \rangle^2\, \frac{ 4 I_1} {1- 16 G_a I_1},
\label{chi4}
\end{equation}
where $I_1= - {\langle \bar q q \rangle}_i /{4 M_i}$.

The present PNJL model has  four parameters in the NJL sector: $m$, $\Lambda$, $g_1$, and
$g_2$. We choose to adjust the parameters in vacuum by fitting to well-known experimental
data or lattice values: the mass of the pion, its decay constant, the quark condensate
and the topological susceptibility that  are shown in Table \ref{table:paramNJL}.
The masses of the $\sigma$, $\eta$, and $a_0$  mesons come as outputs.
This set of parameters is crucial to get the correct description of isentropic trajectories in the
$T\rightarrow 0$ limit \cite{parameters}.

\begin{table}[t]
\begin{center}
\begin{tabular}{||c||c|c|c|c|c|c|c||}
\hline\hline
& $f_\pi$ & $\ave{\bar qq}^{1/3}$  & $m_\pi$ & $m_\sigma$ & $m_\eta$ & $m_{a_0}$ & $\chi^{1/4}$\\
& [MeV]   &  [MeV]                 & [MeV]   & [MeV]      & [MeV]    & [MeV]     & [MeV]\\
\hline
\hline
Model & $93$ & $-241$ & $140.2$ & $803.7$  &  $704.5$ & $919.8$ & $180.8$  \\
\hline
Experimental /Lattice & $92.4$   &  $-267$   &  $135.0$ & $400-1200$ & $547.3$ & $984.7$ & $180$ \\
\hline\hline
\end{tabular}
\end{center}
\caption{Numerical values for the calculated observables compared with experimental and lattice results,
obtained with $\Lambda= 590$ MeV, $G_s\Lambda^2=G\Lambda^2 = 2.435$, $c=0.2$, and $m=6$ MeV.}
\label{table:paramNJL}
\end{table}


\section{Results and discussion}
\label{sec:results}

In a previous work \cite{Santos},  from the analysis of the behavior of the  the topological
susceptibility and of meson axial partners, we concluded that effective restoration of axial
symmetry could only be achieved when $g_2$ was taken as a decreasing function of temperature.
By using $g_2(T) = g_2(0)/\left(1 + \exp ((T - T_0)/10)\right)$ we got the full restoration
of axial symmetry even with a finite cutoff, $\Lambda$, at finite temperature; this ansatz is
interpreted as an explicit mechanism of instanton suppression. In the present work, we will
enlarge the investigation of possible temperature dependence of the coupling coefficients and
its consequences for several observables. We  consider two scenarios, that in the PNJL model are
(see Table \ref{table:couplTemp})
\begin{itemize}
    \item  {\em Scenario {A}} {---}  We will keep $g_1 $ and $g_2 $ as independent parameters.
    At finite temperature we may allow $g_2$ to have an explicit dependence on temperature but $g_1$
    is kept constant.
    \item {\em Scenario {B}} {---} We will use the redefinition of Eq. (\ref{redef}) allowing for
\end{itemize}
		\begin{equation}
    		g_1 = G(1 - c), \,\,\,\,\,\,g_2 = G\,c.\,\,\,\,\,
		\end{equation}
		In the last scenario $g_1$ and $g_2$ are not independent, but $G_s=G$ will be kept always
		constant; on the contrary, $G_a$ varies  since
\begin{equation}
c(T)= 0.2 f(T), \,\,{\rm where} \,\,\, f(T) = 1/\left(1+\exp\left((T -T_0)/10\right)\right).
\label{coefT}
\end{equation}

\begin{table}[t]
\begin{center}
\begin{tabular}{||c||c|c|c||}
\hline\hline
&  I &  II  & III  \\
&&(\textit{Scenario A}) & (\textit{Scenario  B}) \\
\hline
PNJL & $g_1, \, g_2$ & $g_1,\, g_2 (T)$ & $g_1 (T),\, g_2 (T)$ \\
\hline
EPNJL & $g_1 (\Phi),\, g_2(\Phi)$ & $g_1(\Phi),\, g_2 (\Phi, T)$ & $g_1 (\Phi, T),\, g_2 (\Phi, T)$ \\
\hline\hline
\end{tabular}
\end{center}
\caption{Scenarios of the temperature dependence of $g_1 $ and $g_2$ in PNJL and EPNJL models:
in column I there is no explicit dependence on the temperature, columns II and III correspond
to the scenarios A and B.}
\label{table:couplTemp}
\end{table}

In this case only the topological susceptibility and the $\eta$ and $a_0$ meson masses will be affected.
The other quantities have the same behavior as when $g_1$ and $g_2$ are kept constant.

In the EPNJL we have equivalent scenarios with the replacement:
$g_1 \rightarrow g_1(\Phi)$ and $g_2 \rightarrow g_2(\Phi)$.
In both cases we  allow $T_0$ (the critical temperature  for the deconfinement phase
transition within a pure gauge approach, usually fixed to $270$ MeV) to have  several
values and  we discuss this effect. We will compare results obtained in the framework of
the PNJL and EPNJL models for the characteristic temperatures, the pressure, the
topological susceptibility, and the masses of chiral and axial meson partners. Along the
work, we will always consider the cutoff $\Lambda\rightarrow\infty$ at finite
temperature \cite{Regulariz,Rossner:2007PRD}. As explained in Ref. \cite{Regulariz}, 
above the temperature at which the symmetry dynamically broken is fully restored, 
that is  $M_i=m_i$, the condensates are set to zero.  The use of the infinite cutoff allows the 
presence of high momentum quarks, ensuring that the pressure goes to the Stefan-Boltzmann limit and, 
as shown in  \cite{parameters}  gives a better description of several thermodynamic quantities.

As it can be seen in Table IV, in the PNJL model $T_{\chi}$ and $T_d$
never coincide but are closer for higher values of $T_0$, so it is adequate
$T_0 \simeq 270$ MeV, the value used in the pure gauge approach. 
We also present results for $T_0=210$ MeV, the value derived by RG considerations. 
The transition temperatures are defined by the peaks in the susceptibilities of the 
chiral condensate, for the restoration of chiral symmetry, and of the Polyakov loop, 
for the confinement-deconfinement transition.

However, in the EPNJL model, where $\Delta=(T_{\chi}-T_d)/T_{\chi}=0$
 by construction, we have more freedom to fix $T_0$ and a lower value of $T_0$
is convenient $(\simeq 170\mbox{ MeV})$ since it allows one to reproduce lattice results
for the critical temperature for deconfinement and restoration of chiral symmetry  \cite{lattice_Tc}.
The results are presented for scenario B, for reasons that will be explained latter.
\begin{table}[t]
    \begin{center}
        \begin{tabular}{||c||c|c|c|c|c||}
            \hhline{|-|-|-|-|-|-|}
            Scenario B 	& $T_0$ & $T_{\chi}$	& $T_d$	& $\Delta$ 	& $T_{eff}$ \\
            						& [MeV] & [MeV] 			& [MeV]	& 	--			& [MeV] 		\\
          	\hline
            \hline
             PNJL   & 210 & 215 & 177 & 18\% & $\sim$ 250 \\
             $ $   & 270 & 237 & 219 &  8\% & $\sim$ 300\\
            \hline
            EPNJL  & 170 & 173 & 173 & -- & $\sim$ 200\\
             $  $  & 270 & 223 & 223 & -- & $\sim$ 300\\
            \hhline{|-|-|-|-|-|-|}
        \end{tabular}
         \caption{
          Characteristic temperatures in the PNJL and the EPNJL model for different values of
          $T_0$ $\left(\Delta=(T_{\chi}-T_d)/T_{\chi}\right)$.}
    \end{center}
\label{table:Temperatures}
\end{table}

In the EPNJL model the coupling constants are replaced by effective couplings dependent on $\Phi$,
which ensures the entanglement between deconfinement and restoration of chiral symmetry.
Now, using a EPNJL model with anomaly, a question arises: is the restoration of axial symmetry also
entangled with two phase transitions mentioned above or should we have, like in PNJL
model, an independent mechanism of instanton suppression?
Having in mind  the PNJL results \cite{Santos}, we will discuss  results for the following cases that
are summarized in Table \ref{table:couplTemp}:
\begin{itemize}
	\item  G($\Phi$), with $g_1(\Phi)$, $g_2(\Phi)$, without extra dependence on temperature.
	\item  G($\Phi$), with $g_1(\Phi,T)$, $g_ 2(\Phi,T)$, scenario B, the explicit dependence on
	temperature being introduced, as in the PNJL model, through Eq. (\ref{coefT}).
\end{itemize}

\begin{figure}[t]
\begin{center}
  \begin{tabular}{cc}
       \hspace*{-0.5cm}\epsfig{file=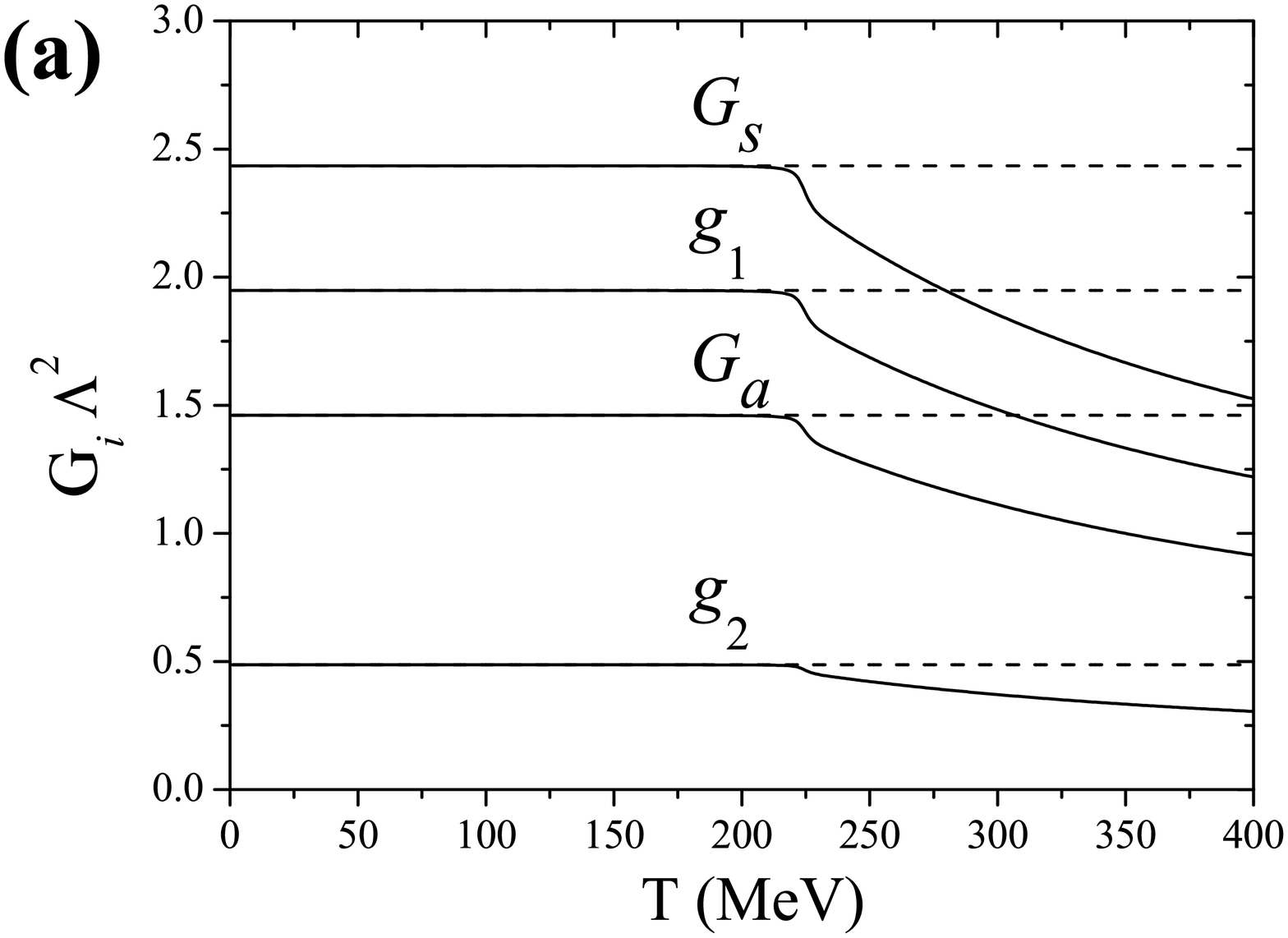,width=8.5cm,height=6.5cm}&
       \hspace*{-0.5cm}\epsfig{file=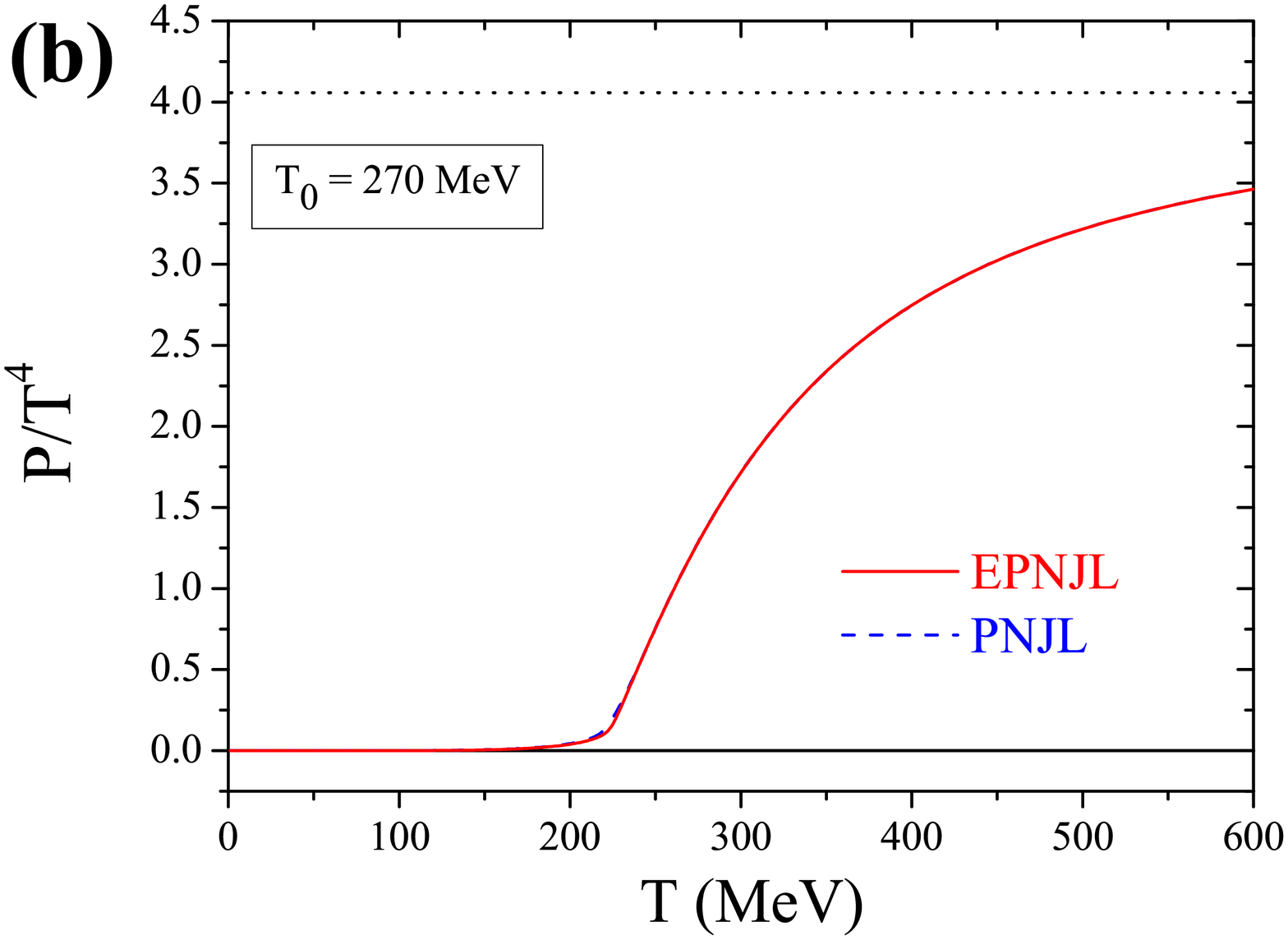,width=8.5cm,height=6.5cm}
  \end{tabular}
    \epsfig{file=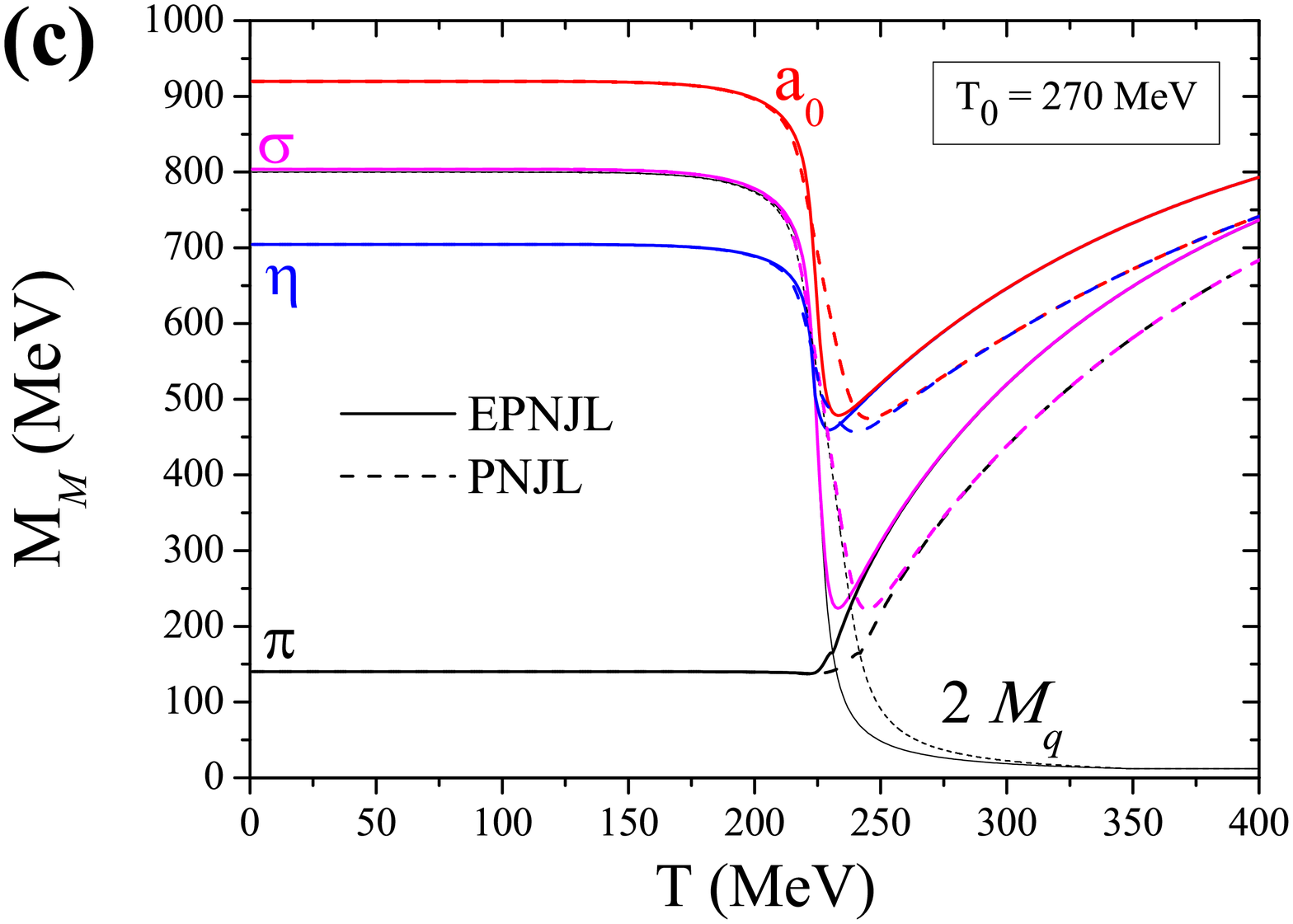,width=8.5cm,height=6.5cm}
\end{center}
\caption {Comparison of several quantities in the PNJL model (dashed lines) and the EPNJL model (full lines),
without explicit dependence of $g_1$ and $g_2$ on temperature and at $T_0=270$ MeV: coupling vertices
(a), pressure (b), and meson masses (c). }
\label{fig:PNJL_EPNJL}
\end{figure}

Let us concentrate on observables related with the restoration of axial symmetry.
Concerning the topological susceptibility, its vanishing is guaranteed in both models
since the infinite cutoff  leads to the vanishing of the quark condensate and $\chi$
is proportional to $\left\langle \bar qq\right\rangle^2$ [see Eq. (\ref{Eq:gap})].
As for the gap between the masses of the meson axial partners, in order to get
its vanishing one should have $G_a \rightarrow G_s$.
We begin by considering  no explicit mechanism of instanton suppression in the models,
that is: we have $g_1$ and $g_2$ constants, in the PNJL, and $g_1(\Phi)$ and $g_2(\Phi)$
in EPNJL [column I of Table III].
As it can be seen in Fig. \ref{fig:PNJL_EPNJL}(a) for $T_0=270$ MeV, although in the EPNJL model
the effective vertices have a temperature dependence  through the Polyakov field, $\Phi$,
$G_s$ does not converge to $G_a$ and, consequently, the axial partners
$(\pi, a_0)$ and $(\sigma, \eta)$ do not degenerate in Fig. (\ref{fig:PNJL_EPNJL}(c)).
Therefore, we conclude that in both models we need an explicit temperature dependence of
the coupling coefficients in order to have effective restoration of axial symmetry.
Finally, we notice that the results for the pressure are qualitatively similar in both models
[Fig. \ref{fig:PNJL_EPNJL}(b)].

\begin{figure}[t]
\begin{center}
  \begin{tabular}{cc}
       \hspace*{-0.5cm}\epsfig{file=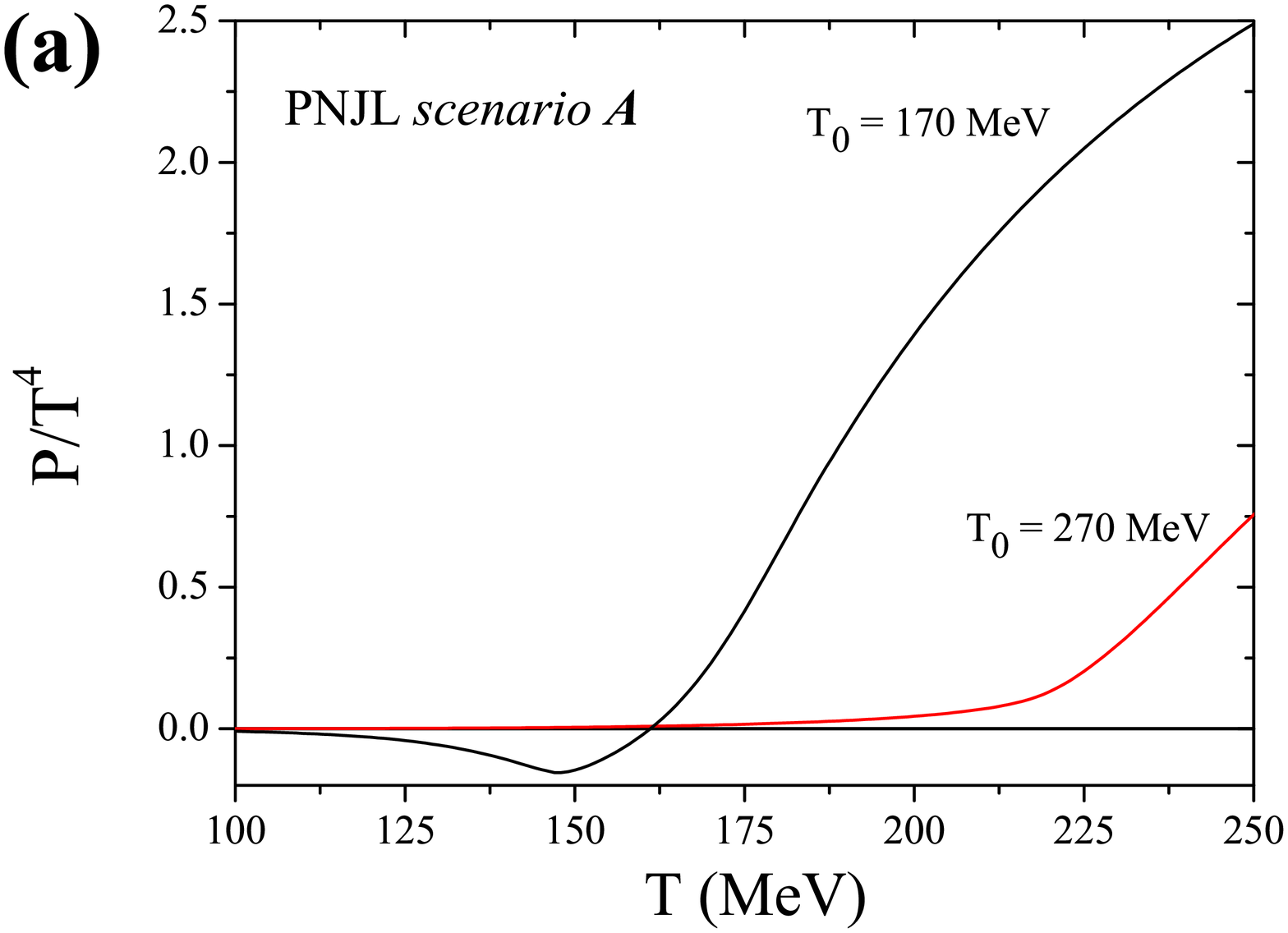,width=8.5cm,height=6.5cm}&
       \hspace*{-0.5cm}\epsfig{file=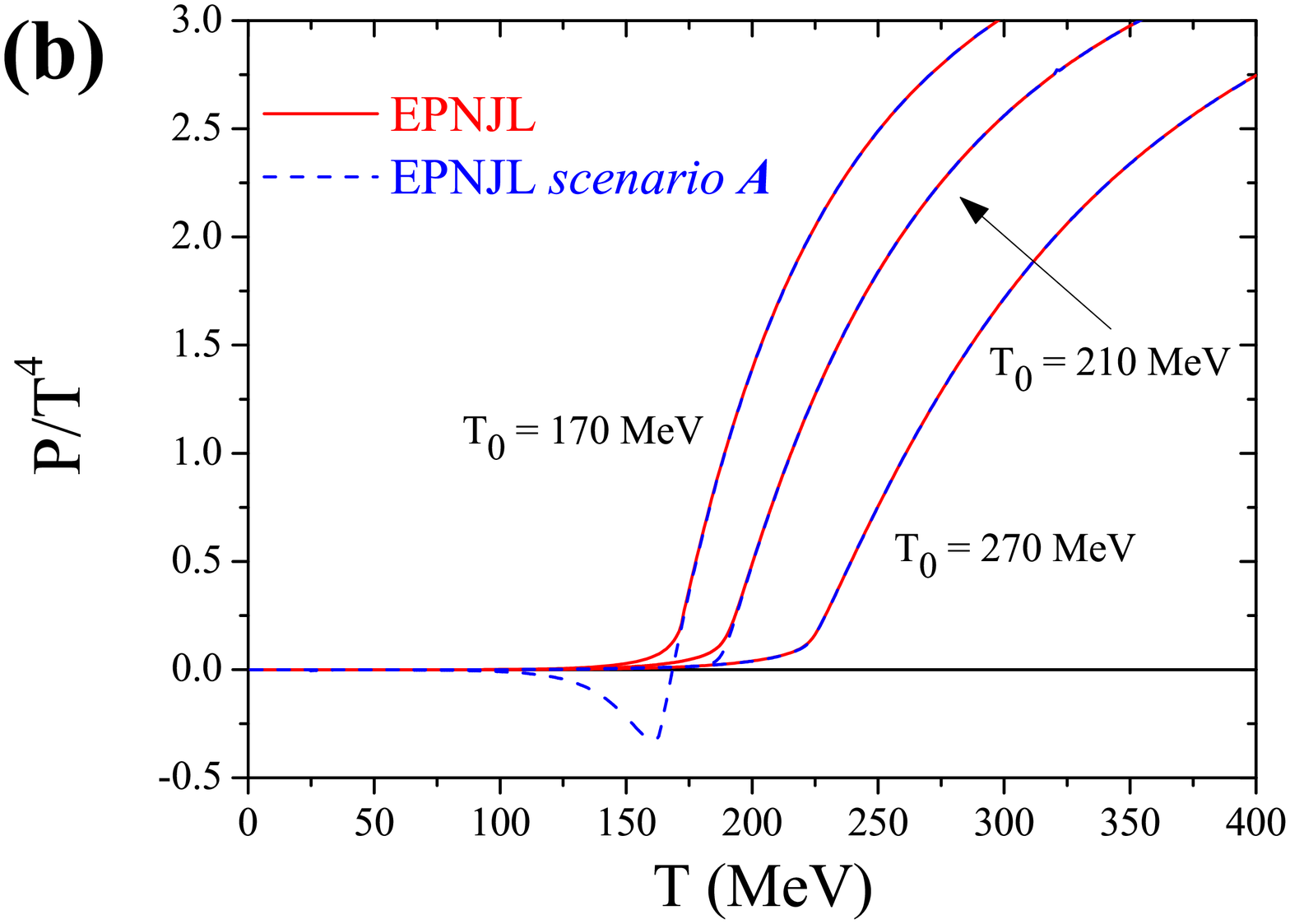,width=8.5cm,height=6.5cm}
  \end{tabular}
    \epsfig{file=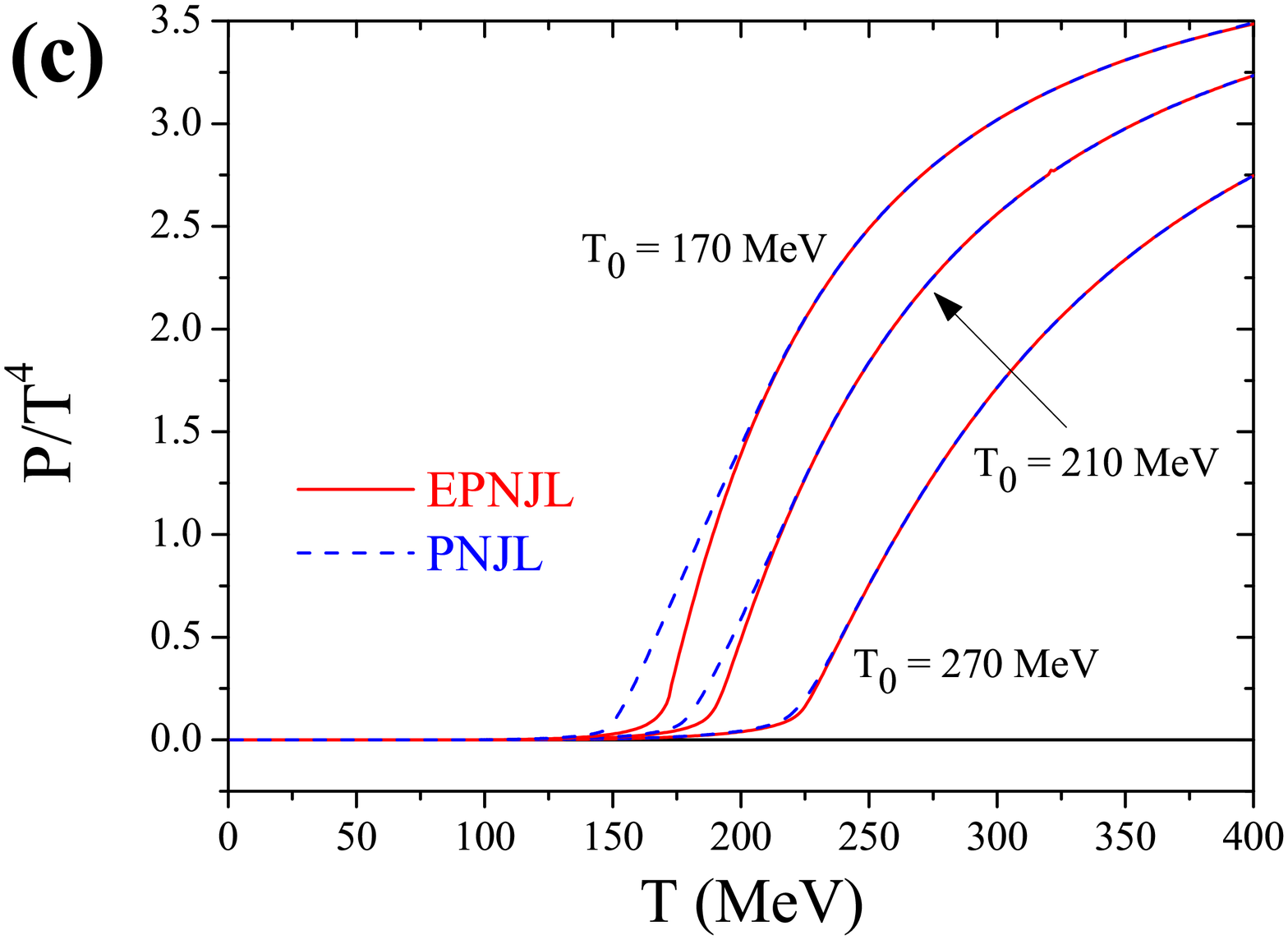,width=8.5cm,height=6.5cm}
\end{center}
\caption {Pressure for different values of $T_0$ with scenario A in PNJL and EPNJL
models [panels (a) and (b)]; with scenario B in PNJL and EPNJL models [panel (c)].}
\label{fig:press_PNJL_EPNJL}
\end{figure}

In order to check  whether the scenarios explored   for the temperature dependence
are physically meaningful, we begin by plotting  in Fig. \ref{fig:press_PNJL_EPNJL}
the pressure  for different scenarios.
We see that in scenario A, for both models,  there is an unphysical region of negative
pressures for low values of $T_0$ [Figs. 2(a) 2(b)].
This is probably due to a too fast instanton suppression when $T_0$ takes lower values,
so we need a mechanism to balance this effect if we want to consider low values of $T_0$.
This problem does not exist in scenario B [Fig. \ref{fig:press_PNJL_EPNJL}(c)], and,
as will be seen (Figs. \ref{fig:PNJL} and \ref{fig:EPNJL}), this ansatz also guarantees the
restoration of axial symmetry. So, scenario B, which allows freedom in fixing $T_0$, will
be  the scenario adopted in the remainder of the present work. 
The difference in the behavior of the pressure in both scenarios can be understood from 
Eq. (\ref{omega}), which depends explicitly on the coupling constants through $G_s=g_1+g_2$. 
The coupling $G_s$ depends on $c(T)$ [Eq. (\ref{coefT})] in scenario A, but not in scenario B.

\begin{figure}[t]
\begin{center}
  \begin{tabular}{cc}
       \hspace*{-0.5cm}\epsfig{file=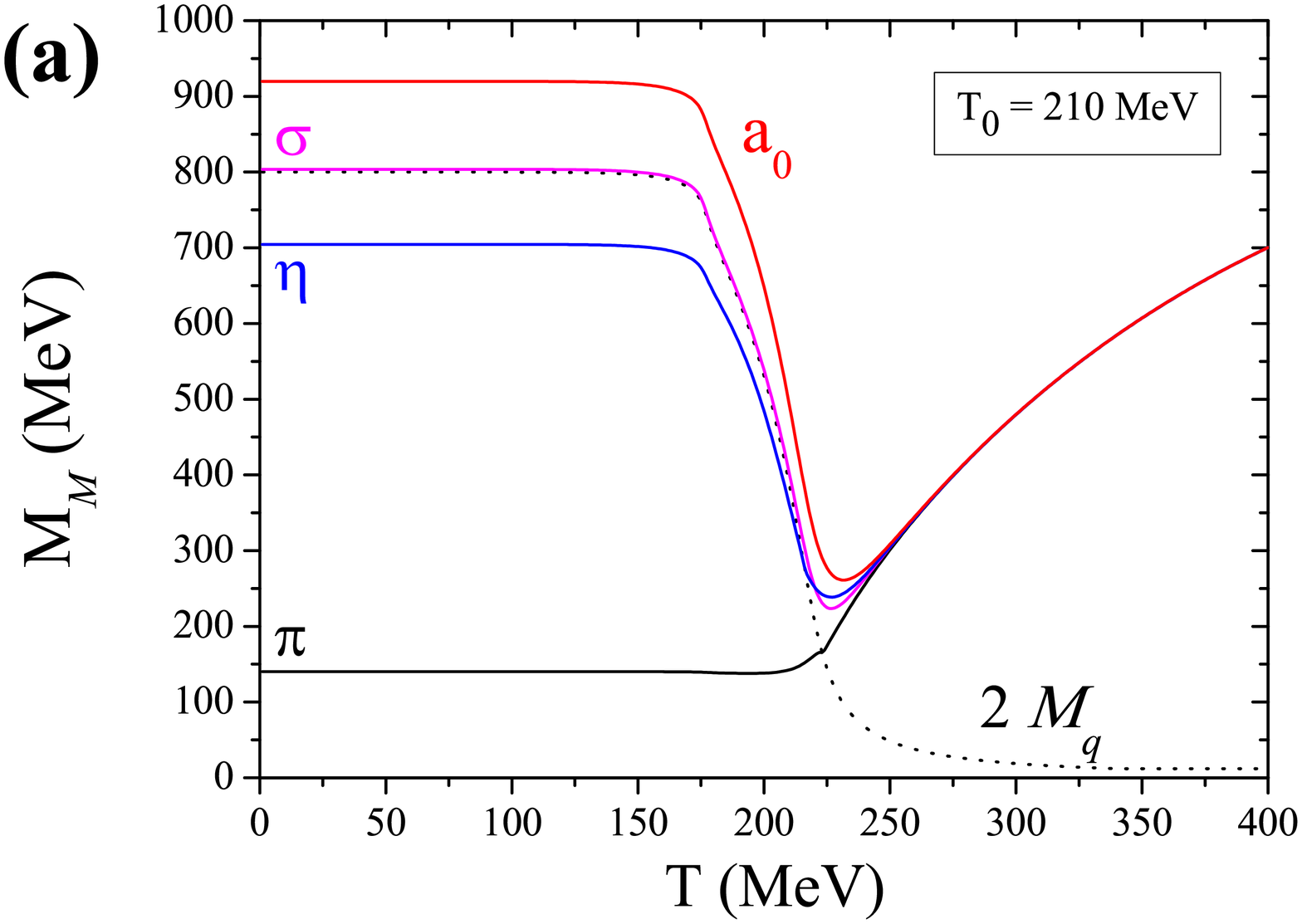,width=8.5cm,height=6.5cm}&
       \hspace*{-0.5cm}\epsfig{file=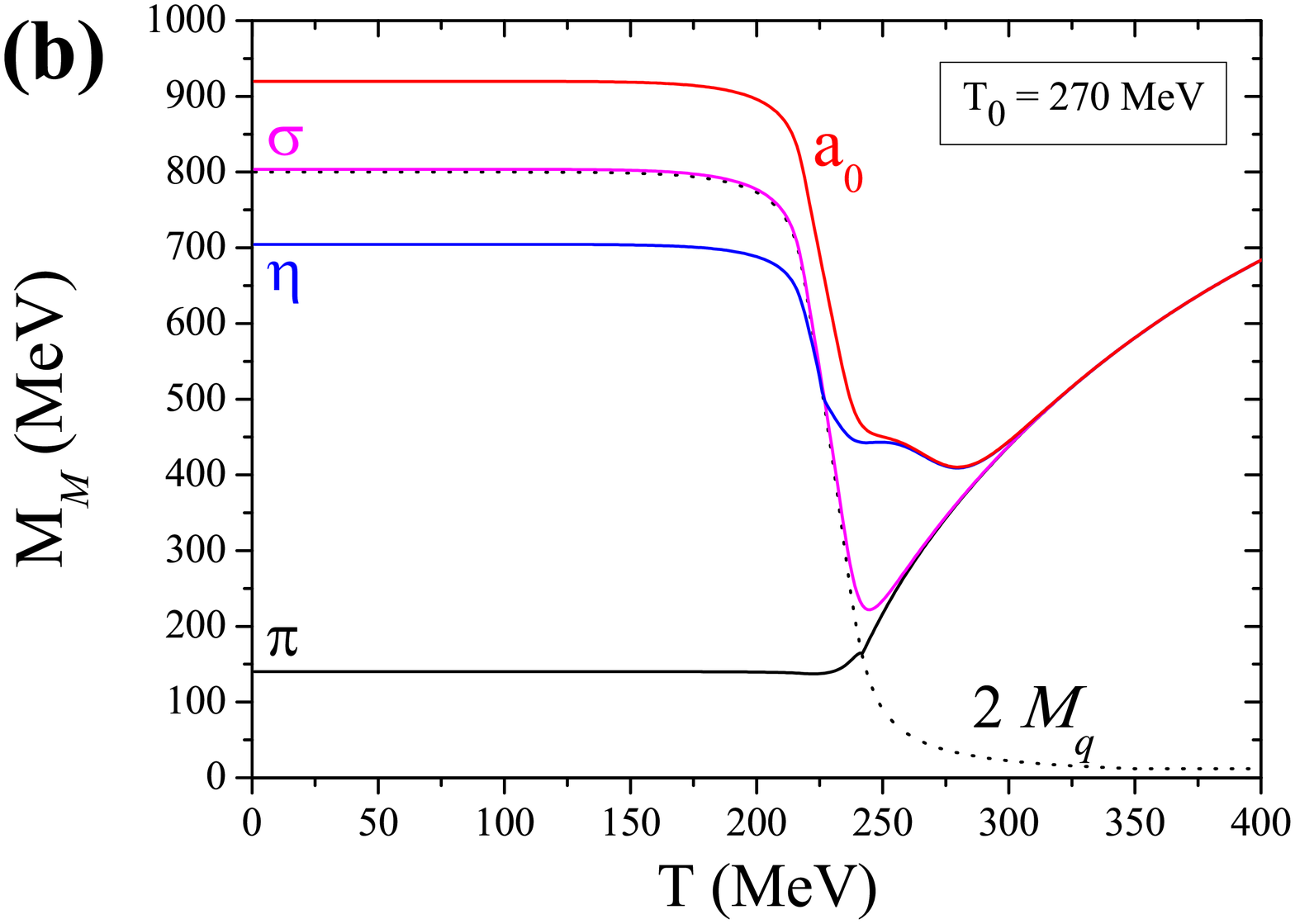,width=8.5cm,height=6.5cm}\\
  \end{tabular}
    \epsfig{file=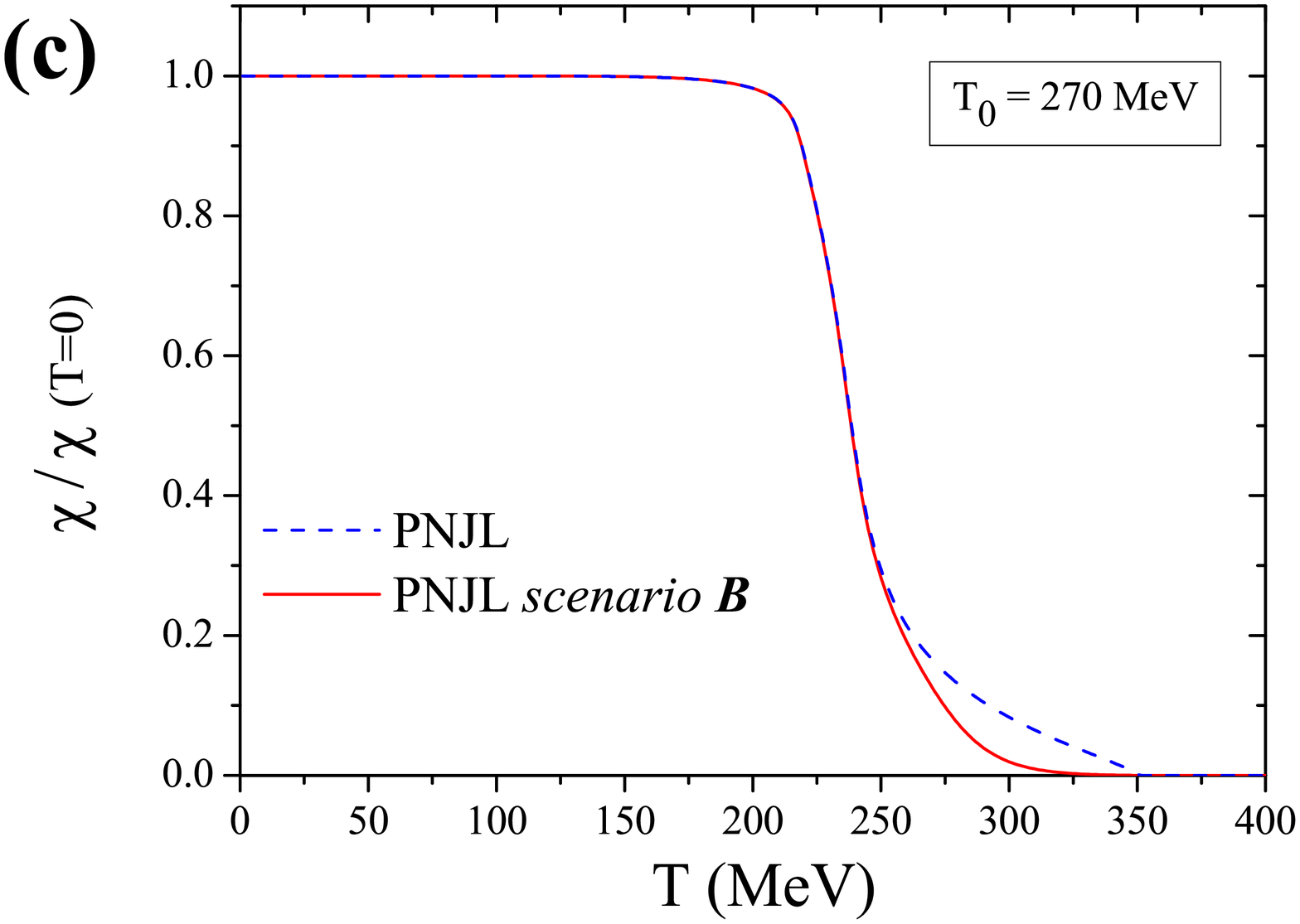,width=8.5cm,height=6.5cm}
\end{center}
\caption { Meson masses in the PNJL model with  scenario B for two values of $T_0$ and
topological susceptibility with different scenarios.}
\label{fig:PNJL}
\end{figure}

In Fig. \ref{fig:PNJL} we plot the PNJL results  for the meson masses,
considering two values of $T_0$, in  scenario B,
and  the topological susceptibility for different cases. We see that both the
topological susceptibility as well as the gap between chiral and axial partners vanish,
so axial symmetry is effectively restored. For $T_0 = 270$ MeV the convergence of chiral
partners occurs before that of axial partners, as usual, with $T_{eff}\approx 300$ MeV
($T_{eff}$ is the temperature at which the effective restoration of the two symmetries is
achieved). The new finding is that for $T_0 = 210$ MeV the  chiral and axial partners
get degenerate very closely and  $T_{eff}$ is lower.   An exploration of this
temperature region could lead to a discussion about the sequence  of restoration of two
symmetries or even of its possible coincidence. Concerning the topological susceptibility 
[Fig. \ref{fig:PNJL}(c)], the behavior is quite similar for scenario A and B and it vanishes at
$T\approx 325$ MeV; when both $g_1$ and $g_2$ are kept constant, the topological susceptibility
vanishes later, in this case this effect is only due to the vanishing of the quark
condensate, a consequence of the full restoration of chiral symmetry.

\begin{figure}[t]
\begin{center}
  \begin{tabular}{cc}
       \hspace*{-0.5cm}\epsfig{file=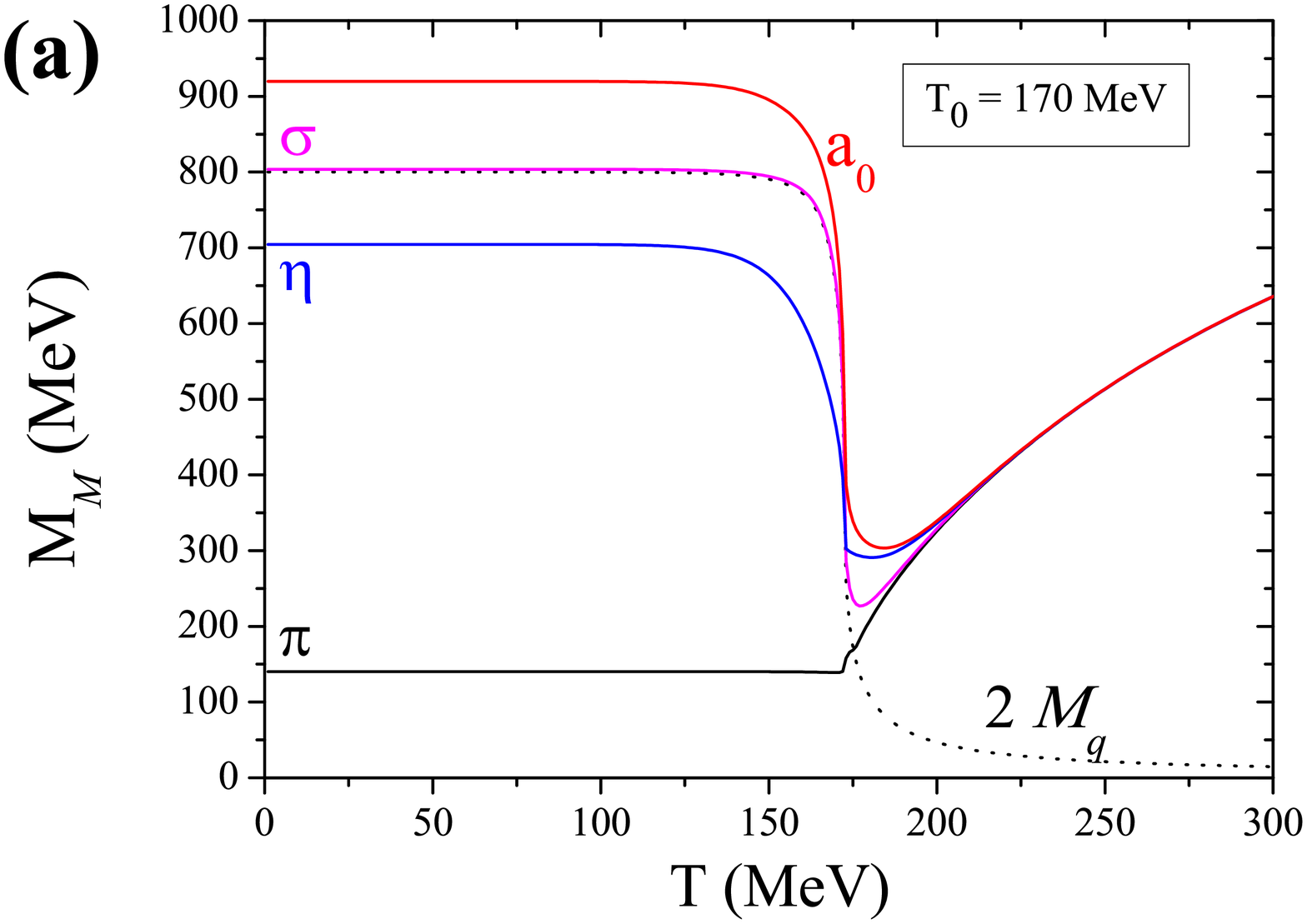,width=8.5cm,height=6.5cm}&
       \hspace*{-0.5cm}\epsfig{file=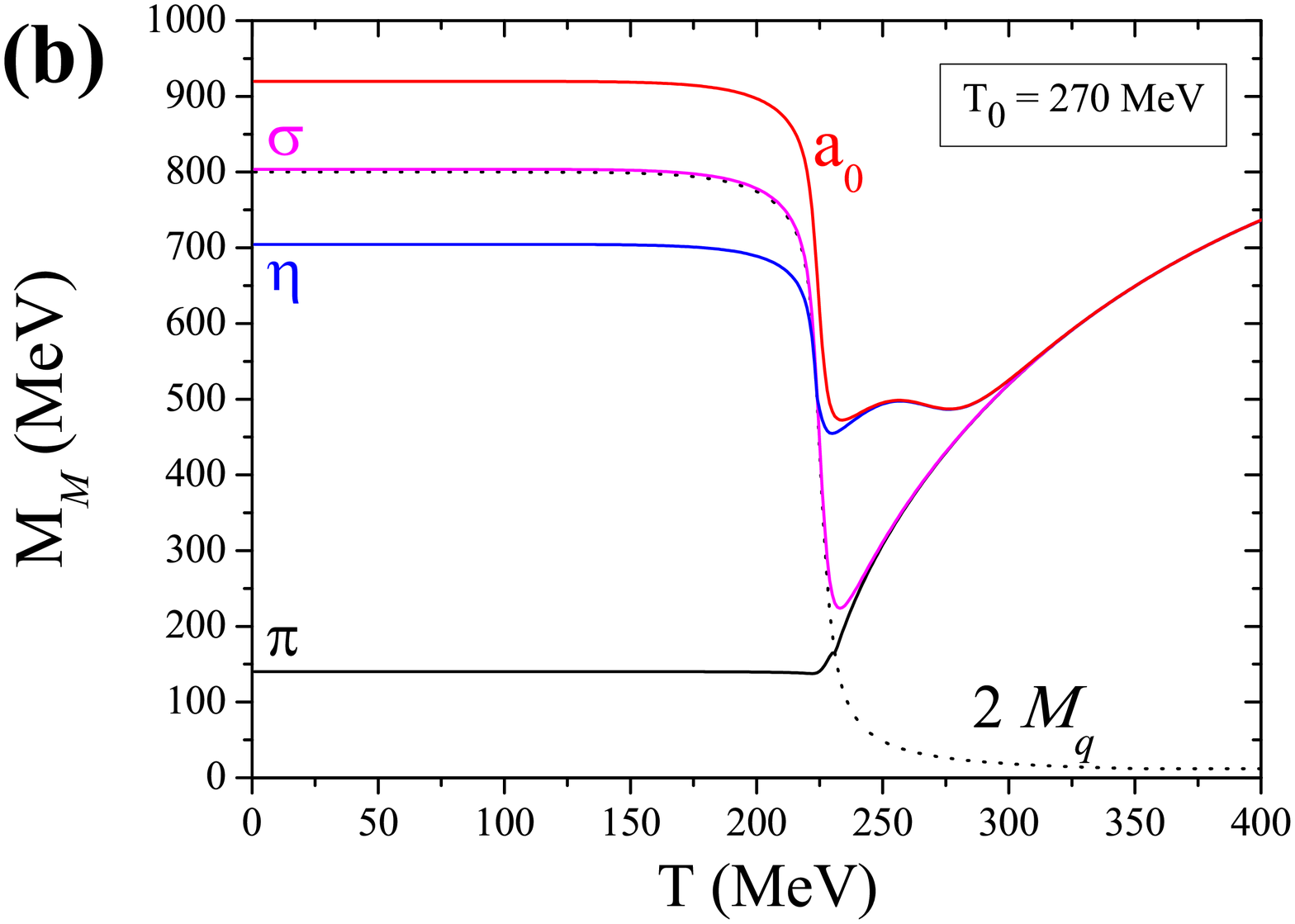,width=8.5cm,height=6.5cm}\\
  \end{tabular}
    \epsfig{file=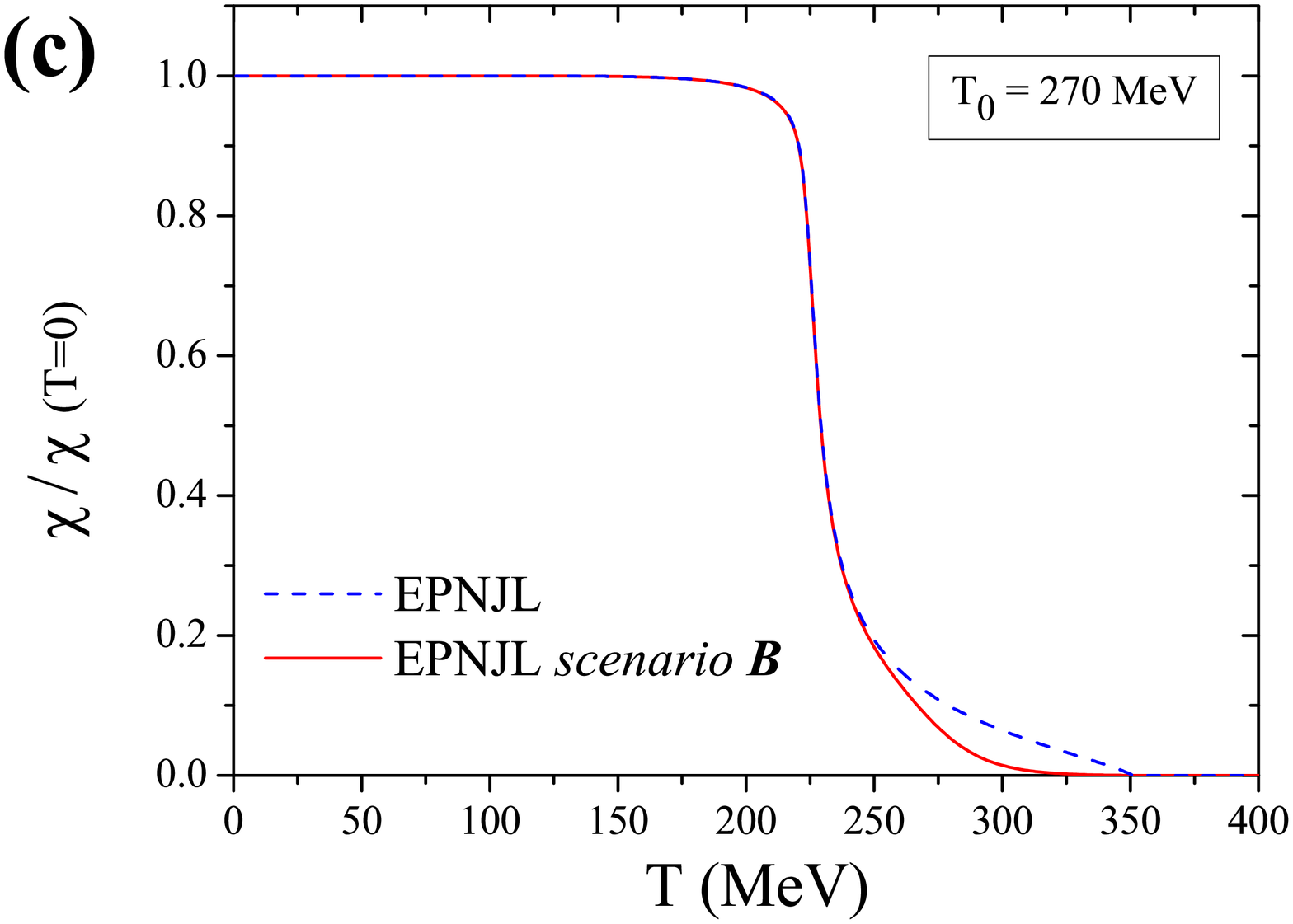,width=8.5cm,height=6.5cm}
\end{center}
\caption {Meson masses in the EPNJL model with  scenario B for two values of $T_0$ [panels (a) and (b)]
and topological susceptibility with two scenarios [panel (c)].}
\label{fig:EPNJL}
\end{figure}

The behavior of the meson masses and topological susceptibility in the EPNJL model is
qualitatively similar to the PNJL model, as it can be seen in Fig. \ref{fig:EPNJL}.
Here, also the restoration of chiral and axial symmetries becomes closer for low values of $T_0$,
but the temperatures for the effective restoration of symmetries are slightly lower.
We conclude that in EPNJL there is entanglement between restoration of chiral symmetry
and deconfinement but not with restoration of axial symmetry.
In order to have restoration of this symmetry
an explicit correlated dependence on $T$ of $g_1 (\Phi)$ and $g_2 (\Phi)$, which ensures instanton
suppression and gives meaningful physical results, is necessary.


\section{Conclusions}

We performed a comparative study of deconfinement,  restoration of   chiral
and axial symmetries within the framework of the PNJL and EPNJL models. In the EPNJL model
the coupling coefficients are endowed with a dependence on the Polyakov field, $\Phi$, which
guarantees that deconfinement and restoration of chiral symmetry occurs at the same
temperature, which can be fitted to lattice result  by choosing a low value for the
parameter $T_0 = 170$ MeV. In order to discuss the effective restoration of chiral
and axial symmetries, we calculate the topological susceptibility and the masses of the
mesons which are chiral and axial partners.
At finite temperature we consider the cutoff $\Lambda \rightarrow \infty$,
in order to get the Stefan-Boltzmann limit for the pressure.

In both models we verified that  the masses of chiral partners $\sigma$ and $\pi$ converge
and that the topological susceptibility vanishes, as a consequence of the effective
restoration of chiral symmetry that occurs when the quark condensate vanishes.
However, if no explicit temperature dependence of the coupling coefficient $g_2$ is considered,
the masses of the axial partners do not converge and anomaly effects remain in the chirally
symmetric vacuum. In order to get the vanishing of the anomaly  it is enough
that only $g_2$ is a decreasing function of the temperature that eventually goes to zero.
However, if there is no mechanism to balance a fast  decreasing of $g_2$, for low values of
$T_0$, an unphysical region of negative pressures occurs. To prevent this, and have more
freedom in the choice of $T_0$, we consider both $g_1$ and $g_2$ as explicit functions of
temperature, $g_2$ going to zero  while $g_1$ attains a  maximum value. This approach also
guarantees the convergence of the axial partners, $(\pi, a_0)$ and $(\sigma, \eta)$.
In both models the restoration of both symmetries is closer when low values of $T_0$ are used.
Our results follow the tendency of recent  lattice results (calculated in the chiral limit),
in particular in what concerns  the degeneracy  of  mesons chiral and axial partners
\cite{UA1JLQCD,UA1lattice,UA1lattice2}.
Finally, we remark that although in the  EPNJL model there is entanglement between restoration
of chiral symmetry and deconfinememt, the restoration of the axial symmetry needs an extra
assumption that ensures the annihilation of the contribution of the 't Hooft interaction.
It should be noticed that the critical temperature, $T_{\chi}$, signals  only the partial
restoration of chiral  symmetry, its effective restoration occurring only at $T_{eff}$ when the
chiral partners get degenerate. So, even when chiral and axial symmetry become effectively
restored, since $T_{eff} > T_{\chi}$,  some effects of the breaking of both symmetries  remain
above the critical temperature, in the interval $( T_{\chi},  T_{eff})$.

\begin{acknowledgments}
This work was supported by Projects No. CERN/FP/116356/2010 and No. PTDC/FIS/\\100968/2008,
projects developed under the initiative QREN financed by the UE/FEDER through
the Program COMPETE - ``Programa Operacional Factores de Competitividade''.
\end{acknowledgments}
\vspace{0.5cm}



\begin{thebibliography}{}

\bibitem{Meisinger:1996PLB}
    P. N. Meisinger, and M. C. Ogilve,
    Phys.  Lett. B {\bf 379}, 163 (1996).

\bibitem{Fukushima:2004PLB}
    K. Fukushima,
    Phys. Lett. B {\bf591}, 277 (2004).

\bibitem{Ratti:2005PRD}
    C. Ratti, M. A. Thaler, and W. Weise,
    Phys. Rev. D {\bf 73}, 014019 (2006).

\bibitem{Megias:2006PRD}
    E. Megias, E. Ruiz Arriola, and L.L. Salcedo,
    Phys. Rev. D {\bf 74}, 065005 (2006);
    Phys. Rev. D {\bf 74}, 114014 (2006).

\bibitem{lattice_Tc3F}
		A. Bazavov  $et$ $al.$ (HotQCD Collaboration),
 		Phys. Rev. D {\bf 85}, 054503(2012).
 		
\bibitem{lattice_Tc}
		V. Bornyakov $et$ $al.$ (QCDSF-DIK Collaboration),
 		Phys. Rev. D {\bf 82}, 014504 (2010);
 		arXiv:1102.4461.

\bibitem{Sakai1}
		Y. Sakai, T. Sasaki, H. Kouno, and M. Yahiro,
    Phys. Rev. D {\bf 82}, 076003 (2010);
    J. Phys. G: Nucl. Part. Phys. {\bf 39}, 035004 (2012).

\bibitem{Sakai2}
 		H. Kouno, Y. Sakai, T. Sasaki, K. Kashiwa, and M. Yahiro,
    Phys. Rev. D {\bf 83}, 076009 (2011).

\bibitem{UA1lattice}
		A. Bazavov, $et$ $al.$ (HotQCD Collaboration),
		Phys. Rev. D {\bf 86}, 094503 (2012).
		
\bibitem{UA1lattice2}
		S. Aoki, H. Fukaya, Y. Taniguchi
		arXiv:1209.2061 [hep-lat].

\bibitem{Alles}
		B. Alles, M. D'Elia, and A. Di Giacomo,
		Nucl. Phys. \textbf{B494}, 281 (1997);
		B. Alles and M. D'Elia,
		arXiv:hep-lat/0602032.

\bibitem{UA1JLQCD}
		G. Cossu $et$ $al.$ (JLQCD Collaboration),
		Proc. Sci., LATTICE2011, 188 (2011),
		[arXiv:1204.4519 [hep-lat]].
		
\bibitem{kunihiro}
		T. Kunihiro,
		Phys. Lett. B {\bf 219} 363(1989);
 		T. Hatsuda and T Kunihiro,
 		Phys. Rep. {\bf 247}, 221 (1994).

\bibitem{Costa1}
		P. Costa, M. C. Ruivo and Yu.L. Kalinovsky,
		Phys. Lett. B {\bf 560,} 171(2003);
		P. Costa, M. C. Ruivo, C.A. de Sousa, and Yu.L. Kalinovsky,
		Phys.Rev. C {\bf 70}, 025204 (2004).
		
		
\bibitem{Ruivo}
		P. Costa, M. C. Ruivo, C.A. de Sousa, and Yu.L. Kalinovsky,
		Phys. Rev. D {\bf 70}, 116013 (2004);
		Phys. Rev. D {\bf 71}, 116002 (2005).
		
\bibitem{veneziano}
		E. Witten,
		Nucl. Phys. B {\bf 156}, 269 (1979);
		G. Veneziano,
		Nucl. Phys. {\bf 159}, 213 (1979).

\bibitem{eta}
		T. Gs\"{o}rg\"{o}, R. V\'{e}rtesi and J. Sziklai,
	 	Phys. Rev. Lett.  {\bf 105}, 182301 (2010).
	
\bibitem{chi1}
    K. Fukushima, K.Ohnishi, and K. Ohta,
    Phys. Rev. C \textbf{63}, 045203 (2001).

\bibitem{Costa}
    P. Costa, M. C. Ruivo, C. A. de Sousa, H. Hansen, and W. M. Alberico,
    Phys. Rev. D \textbf{79}, 116003 (2009);
    P. Costa, M. C. Ruivo, C. A. de Sousa, and H. Hansen,
		Europhys. Lett. \textbf{86}, 31001 (2009);
    Symmetry \textbf{2(3)}, 1338 (2010).

\bibitem{gatto}
    Y. Sakai, H. Kouno, T. Sasaki, and M. Yahiro,
    Phys. Lett. B \textbf{705}, 349 (2011);
    R. Gatto and M. Ruggieri,
    Phys. Rev. D {\bf 85}, 054013 (2012).

\bibitem{Regulariz}
 		P. Costa, M. C. Ruivo, and C. A. de Sousa,
 		Phys.Rev.D {\bf 77}, 096009 (2008).

\bibitem{Santos}
		M. C. Ruivo, M. Santos,  P. Costa, and C. A. de Sousa,
    Phys. Rev. D {\bf 85}, 036001 (2012).

\bibitem{Pisa1}
    R. D. Pisarski,
    Phys. Rev. D {\bf 62}, 111501(R) (2000);
    R. D. Pisarski,
    {arXiv:hep-ph/0203271}.

\bibitem{Kaczmarek:2002mc}
		O.  Kaczmarek,  F.  Karsch,  P.  Petreczky,  and  F.  Zantow,
		Phys.  Lett.  B {\bf 543}, 41 (2002).

\bibitem{Kaczmarek:2007pb}
		O. Kaczmarek,
		Proc. Sci., CPOD07, 043 (2007), [arXiv:0710.0498].
		
\bibitem{Hansen:2007PRD}
		H. Hansen, W. M. Alberico, A. Beraudo, A. Molinari, M. Nardi, and C. Ratti,
    Phys. Rev. D {\bf 75}, 065004 (2007).

\bibitem{Rossner:2007PRD}
		C. Ratti, S. R\"{o}ssner, M. A. Thaler and W. Weise,
	 	Eur. Phys. J. C {\bf 49}, 213 (2007);
		S. R\"{o}ssner, C. Ratti and W. Weise,
	 	Phys. Rev. D {\bf 75}, 034007 (2007).

\bibitem{buballa2}
    M. Frank, M. Buballa, and M. Oertel,
    Phys. Lett. B \textbf{ 562}, 221 (2003).

\bibitem{Brauner:2009gu}
    T. Brauner, K. Fukushima, and Y. Hidaka,
    Phys. Rev. D {\bf 80}, 074035 (2009);
    {\bf 81}, 119904(E) (2010).

\bibitem{Schaefer}
		B.-J. Schaefer, J. M. Pawlowski, and J. Wambach
   	Phys. Rev. D {\bf 76},  074023 (2007).
	
\bibitem{Philipsen}
		O. Philipsen,
		arXiv:1207.5999  [hep-lat].
		
\bibitem{parameters}
		P. Costa, H. Hansen, M. C. Ruivo, and C. A. de Sousa
    Phys. Rev. D {\bf 81}, 016007 (2010);
		M. C. Ruivo, P. Costa, C. A de Sousa, H. Hansen and W. Alberico,
 		AIP Conf. Proc. 1257, 770 (2010)[arXiv:1001.3072 [hep-ph]].
		
\end{thebibliography}
\end{document}